\documentclass[showpacs,twocolumn]{revtex4}
\usepackage{graphicx}
\usepackage{amsmath}
\usepackage{amssymb}
\usepackage{multirow}
\usepackage{color}
\usepackage{float}
\usepackage{lipsum}

\usepackage{}
\begin{document}
\title{Diffractive molecular orbital tomography}
\author{Chunyang Zhai,$^{1}$ Xiaosong Zhu,$^{1}$ Pengfei Lan,$^{1,}$\footnote{pengfeilan@hust.edu.cn} Feng Wang,$^{1}$ Lixin He,$^{1}$ Wenjing Shi,$^{1}$ Yang Li,$^{1}$ Min Li,$^{1}$ Qingbin Zhang,$^{1}$ and Peixiang Lu$^{1,2}$\footnote{lupeixiang@hust.edu.cn}}

\affiliation{$^1$  Wuhan National Laboratory for Optoelectronics and School of Physics, \\
Huazhong University of Science and Technology, Wuhan 430074, China\\
$^2$ Laboratory of Optical Information Technology, Wuhan Institute of Technology, Wuhan 430205, China}
\date{\today}
\begin{abstract}
High harmonic generation in the interaction of femtosecond lasers with atoms and molecules opens the path to molecular orbital tomography and to probe the electronic dynamics with attosecond-{\AA}ngstr\"{o}m resolutions. Molecular orbital tomography requires both the amplitude and phase of the high harmonics. Yet the measurement of phases requires sophisticated techniques and represents formidable challenges at present. Here we report a novel scheme, called diffractive molecular orbital tomography, to retrieve the molecular orbital solely from the amplitude of high harmonics without measuring any phase information. We have applied this method to image the molecular orbitals of N$_2$, CO$_2$ and C$_2$H$_2$. The retrieved orbital is further improved by taking account the correction of Coulomb potential. The diffractive molecular orbital tomography scheme, removing the roadblock of phase measurement, significantly simplifies the molecular orbital tomography procedure and paves an efficient and robust way to the imaging of more complex molecules.
\end{abstract}  \maketitle
\noindent \section{Introduction}
Imaging the molecular orbital gives an intuitionistic insight of the molecular structure and provides the opportunities for revealing and understanding the molecular dynamics and chemical reactions. In the past decades, great strides have been taken to develop the imaging methods with ultrashort X-ray pulses from synchrotrons and free electron lasers \cite{r1,r2,r3} and via diffraction of electron pulses \cite{r4,r5}. Even through a good spatial resolution can be achieved \cite{r6}, the temporal resolution is limited to several tens or even hundreds fs \cite{r6,r7} at present. In recent years, an alternative way has also been developed based on the laser-induced recollision process \cite{r8,r9,rc2h2}. It enables one to image the molecular orbital by measuring the high harmonics generated in the interaction of femtosecond laser and molecules, which is called molecular orbital tomography (MOT) \cite{r8}. The most fascinating perspective of this approach is the potential to get the real-time evolution of the molecular orbital, i.e., a molecular movie, with unprecedented attosecond-{\AA}ngstr\"{o}m resolutions. Since the pioneering demonstration of MOT, paramount interests and continuous efforts have been inspired to probe the electronic and molecular dynamics via high harmonic generation (HHG) \cite{r10,r11,r12,r13,r14,r15}.

To access the intriguing goal of real-time molecular movie, one crucial issue is to effectively capture the snapshot of molecular orbitals. There are several roadblocks at present. The problem stems from that MOT requires both the amplitude and phase information of the high harmonics. The phase is a function of both the harmonic order (i.e., the photon energy) and the angle between the molecular axis and the polarization of the driving laser field. However, it represents a formidable challenge to measure the full phase map as a function of the harmonic order and molecular alignment angle. The other challenges lie in the complexity of HHG, such as multielectron effects \cite{r15,r16} and the influence of Coulomb potential\cite{r17,r18,r19,r20}. In recent years, continuous efforts to measure the harmonic phase have been exerted by many groups and several schemes are developed \cite{r10,r15,r21,r22,r23,r24,r25,r26,r27,r28}. Nevertheless these methods either determine the phase as a function of molecular alignment angle but leave the harmonic-order dependence undetermined, or vice versa \cite{r29}. A complete phase map is not accessible until recently by combining the mixed gases and two source interference schemes \cite{r29}, which however requires sophisticated instrumentation. In the previous MOT experiments of N$_2$ and CO$_2$ molecules, the phases are partially or even fully based on the theoretical simulations or assumptions \cite{r8,r10}.

In this work, we circumvent the above roadblocks of MOT by pursuing a novel way to retrieve the molecular orbital solely from the amplitudes of high harmonics without measuring the phase information. Our scheme is based on the oversampling of the high harmonic amplitude, which contains abundant information and enables to reconstruct the molecular orbital without measuring the phase. Such a scheme is analogy to the coherent diffractive imaging \cite{rdiff,rmiao} and we therefore call it diffractive MOT (DMOT). We have experimentally demonstrated the DMOT scheme by retrieving the molecular orbital of N$_2$ with the harmonic spectra alone. The orbital reconstruction is further improved by taking account of the correction of the Coulomb potential. This DMOT method is robust and straightforward for exploiting other molecules, which is demonstrated for CO$_2$ and C$_2$H$_2$ molecules. The multielectron effects are also addressed. Since the phase measurement is not necessary in DMOT, our method significantly simplifies the reconstruction of the molecular orbital, opening an easier and efficient pathway to MOT.
\noindent \section{Method and Experiment}
\subsection{Principle of DMOT}
\begin{figure}[!t]
\centerline{
\includegraphics[width=8cm]{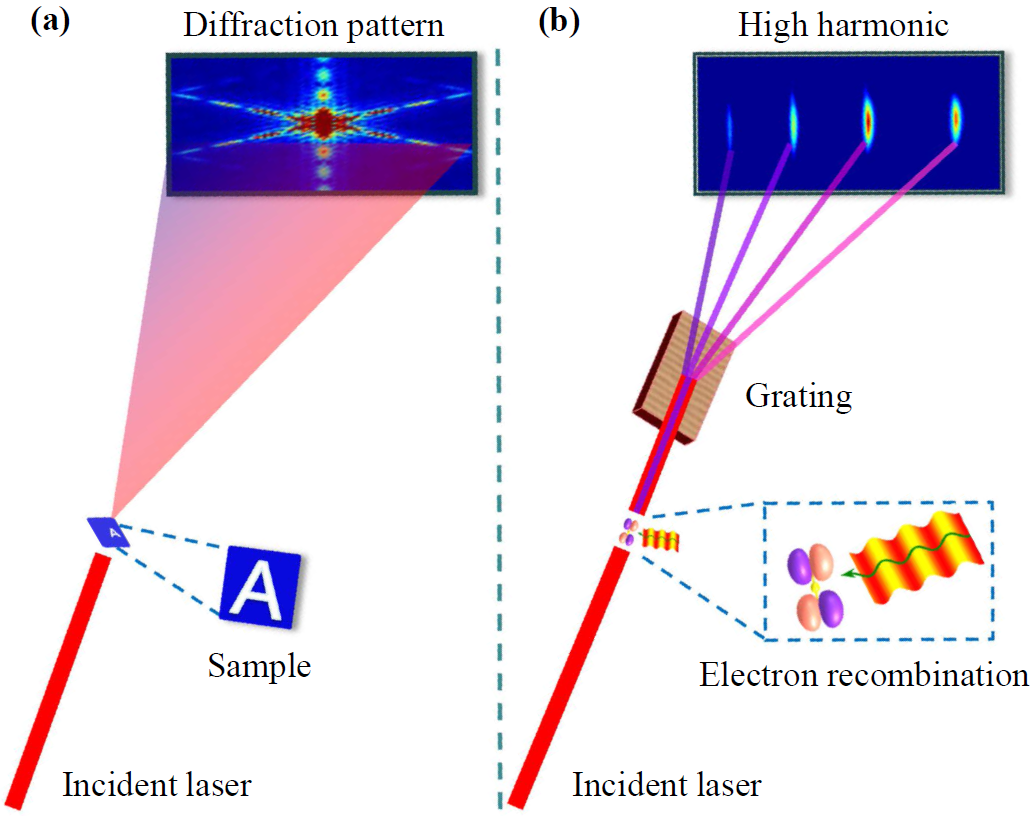}}
\caption{\label{diffractive}(Color online) Illustration of Fraunhofer diffraction and HHG. (a), A coherent wave illustrates on the sample ``A" and the diffraction pattern can be detected in the far field, which only contains the intensity of the diffractive light. (b), Illustration of high harmonic generation by focusing the laser on CO$_2$ molecules. A fractional of the electron wavepacket is freed and accelerated by the laser field and finally returns to the ground state by emitting high harmonics. The intensities of the high harmonics are measured by the grating spectrometer.}
\end{figure}

Figure \ref{diffractive} illustrates the principle of DMOT scheme. According to the three-step model, HHG can be understood by the laser-induced recollision process \cite{r32,r33}, i.e., the most active electron is first freed by tunneling ionization and is then accelerated in the laser field. Finally, the freed electron returns along the laser polarization direction and recombines with the parent ion with emitting high harmonics (see Fig. \ref{diffractive}(b)). The HHG signal $\tilde{\mathbf{E}}(\omega,\theta)$ is proportional to recombination transition dipole $\tilde{\mathbf{d}}(\omega,\theta)$ and can be written as \cite{r8,r34,r47}
\begin{equation} \label{eq1}
\tilde{\mathbf{E}}(\omega,\theta)\propto \tilde{\mathbf{d}}(\omega,\theta)=\mathbf {k}\langle\Psi_0|\mathbf{k}\rangle=\mathbf {k}\mathcal{F}[\Psi_0],
\end{equation}
which is related to the Fourier transform of the molecular orbital $\Psi_0$. Here, the velocity form of recombination dipole is adopted and the notations with upper tilde denote complex values and those without tilde denote real values. $\omega$ in Eq. (\ref{eq1}) is the angular frequency of the high harmonics and $\mathbf{k}$ is the momentum of the returning electron. The Fourier transform relation in Eq. (\ref{eq1}) is analogies to the Fraunhofer diffraction (see Fig. \ref{diffractive}(a)), where a coherent wave illuminates on an object $u(x,y)$ and diffraction pattern in the far field is the Fourier transform of the object, i.e., $\tilde{U}(k_x,k_y)=\mathcal{F}[u(x,y)]$. However, the diffraction patterns measured by the usual detectors, e.g., films or CCD cameras, only contain the information of the amplitude (i.e., $|\tilde{U}(k_x,k_y)|^2$ or $|\tilde{E}(\omega,\theta)|^2$ in HHG) and the phase information is lost. According to the Nyquist-Shannon sampling theorem, the sufficient sampling rate of a Fourier transform signal is the so-called Nyquist rate \cite{r35}. If the diffraction intensity of the object is sampled at spacing finer than half of the Nyquist interval, i.e., oversampling of the diffraction pattern, a nontrivial unique phase set is in principle encoded in the oversampled diffraction pattern \cite{r36}. Then the object can be retrieved from the diffraction intensities with the iterative algorithm \cite{rdiff,rmiao}.

We can show that the equally distributed sampling with the odd high harmonics is several times finer than the Nyquist interval. To this end, we firstly evaluate the Nyquist interval for imaging the molecular orbitals. Since the molecular orbital is a bound state, the wave function decreases exponentially as increasing the distance from the nucleus. Typically, the orbital $|\Psi_0(\mathbf{r})|^2$ is less than 10$^-$$^2$ if $|\mathbf{r}|>5$ a.u.. Therefore, the orbital can be assumed to be a real value function confined within 5 a.u. near the nucleus for the molecules considered in this work. The Nyquist interval in the momentum domain is $2\pi$ divided by the orbital size, i.e., $\bigtriangleup{k_N}=2\pi/10=0.628$ a.u.. To image the molecular orbital, the required sampling interval is smaller than $\bigtriangleup{k_N}/2=0.314$ a.u.. For high harmonic spectra, the frequency interval between the adjacent odd harmonics is $\bigtriangleup\omega=0.114$ a.u. for the 800-nm driving laser. According to the three-step model, the photon energy of high harmonics is determined by the kinetic energy of the returning electron. We have the relation $\omega_q=k_q^2/2$, where $\omega_q$ is the frequency of the \emph{q}th order harmonic and $k_q$ is the momentum of the returning electron. One can obtain that $\bigtriangleup{k}=\bigtriangleup{\omega}/k_q$. Therefore, $\bigtriangleup{k}$ increases as decreasing the harmonic order. It is calculated to be 0.19 a.u. for the third order harmonic (H3). Thus the maximum sampling interval is much smaller than $\bigtriangleup{k_N}/2$. Note that the lowest measured harmonic in most HHG experiment is higher than H3, e.g., the 15th harmonic in our experiment. The maximum sampling interval is $\bigtriangleup{k_{max}}=0.087$ a.u., which is about one quarter of $\bigtriangleup{k_N}/2$. In other words, the harmonic spectra are highly oversampled. The phase information is in principle encoded in the oversampled high harmonic spectra, enabling one to reconstruct the molecular orbital with the high harmonic intensity without measuring the phase as the diffractive imaging. In fact, a sampling interval of 0.087 a.u. in principle supports the imaging of the molecular orbital with the size of 36 a.u., which is large enough for most molecules.

The difference between HHG and diffraction is that the electron recollides with the molecule in only one direction when the molecule is aligned at a specific angle. In other words, the recorded HHG spectrum at one alignment angle is a slice of the diffraction pattern. To image the molecular orbital, one needs to record and assemble the harmonic spectra by varying the alignment angles.
\subsection{Experimental setup}
The experiment is carried out by the pump-probe scheme. Firstly, the molecule is aligned in the laboratory frame by impulsive alignment method \cite{r38} using a non-ionizing and slightly stretched pump pulse. Then the HHG is produced by a time delayed probe pulse. Both the pump and probe pulses are linearly polarized. A 30-fs Ti:sapphire laser with a central wavelength of 800-nm and a 1 kHz repetition rate is used. The maximum energy is 10 mJ per pulse. The laser beam is split and recombined for pump-probe experiment. The pump pulse is stretched to about 50 fs to impulsively align the molecules.The pump and probe pulses are focused on a gas jet with a 600-mm focal-length lens. The nozzle has an orifice of 100 $\mu$m in diameter and the stagnation pressure is 2.3 bars. The gas jet is placed 2 mm downstream of the laser focus to realize the phase-matching of the short quantum trajectory of HHG. The pulse energy is continuously adjusted with a half-wave plate and a polarizer. The beam sizes of the pump and probe pulses are adjusted independently by two diaphragms. The polarization of the pump pulse is changed with a half-wave plate. The high harmonic spectrum is measured by a flat-field soft X-ray spectrometer \cite{r39}.
\noindent \section{Results and Discussions}
\begin{figure}[!t]
\centerline{
\includegraphics[width=8cm]{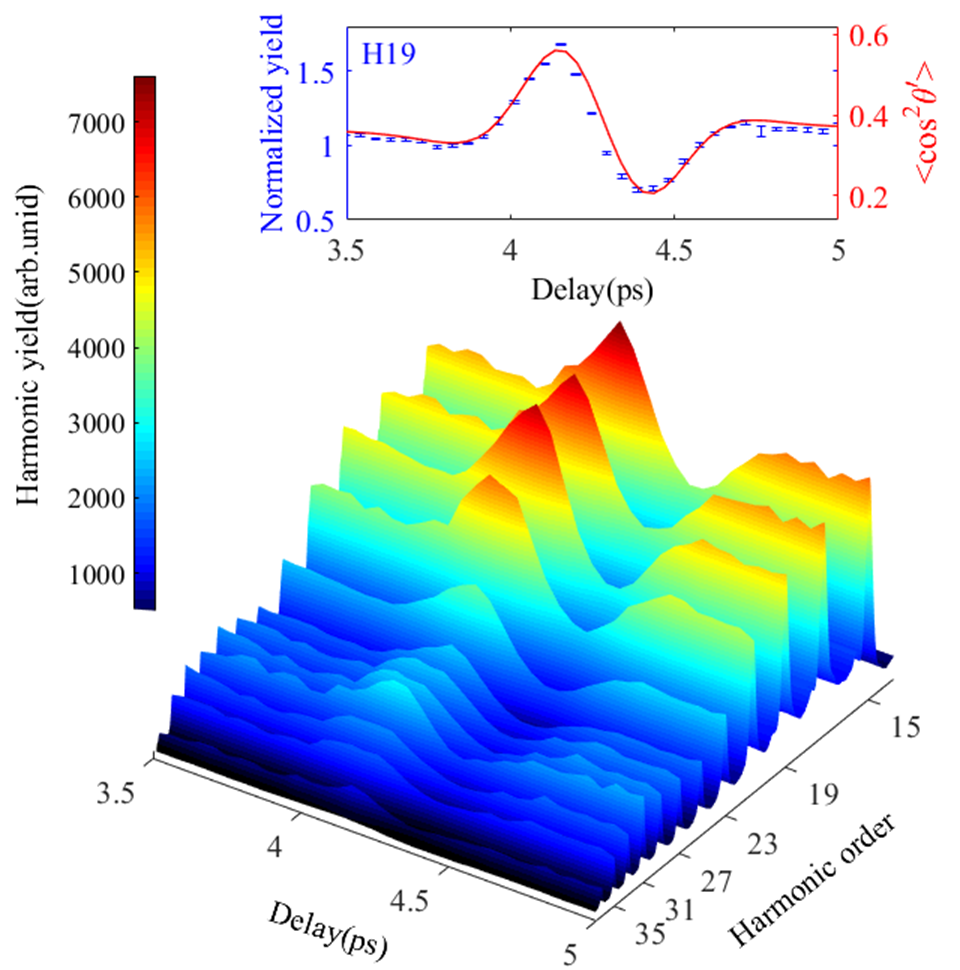}}
\caption{\label{n2delay}(Color online) High harmonic spectra of N$_2$ molecules measured as a function of the pump-probe delay. The pump and probe pulses have parallel polarizations. The solid line in the inset displays the calculated temporal evolution of the alignment degree $\langle\cos^2{\theta}^\prime\rangle$. The dots display normalized HHG yield of the 19th harmonic.}
\end{figure}

We first demonstrate the DMOT scheme with N$_2$ molecules.
The probe pulse intensity is estimated to be $2.4\times10^{14}$ W/cm$^2$. The pump pulse intensity is estimated to be $5\times10^{13}$ W/cm$^2$. First, the polarization of the pump pulse is kept parallel to the probe pulse and high harmonic spectrum is measured by varying the delay between the pump and probe pulses.
Figure \ref{n2delay} shows the harmonic spectra of N$_2$ as scanning the pump-probe delay. One can clearly observe a peak at the delay around 4.1 ps and a valley at 4.4 ps, which corresponds to the quantum half revival of the molecular rotation of N$_2$. The evolution of the alignment degree can be evaluated by $\langle\cos^2{\theta}^\prime\rangle$ \cite{r38,rali}, where ${\theta}^\prime$ is the angle between the molecular axis and the polarization of pump pulse. The simulated result is shown in the inset by red solid curve. The blue dots in the inset show the delay dependence of the 19th harmonic yield, which is normalized to the harmonic yield without pump pulse. One can see that $\langle\cos^2{\theta}^\prime\rangle$ has the same trend with the normalized yield and reaches its maximum at 4.1 ps, where the molecules are well aligned.
\begin{figure}[!b]
\centerline{
\includegraphics[width=8cm]{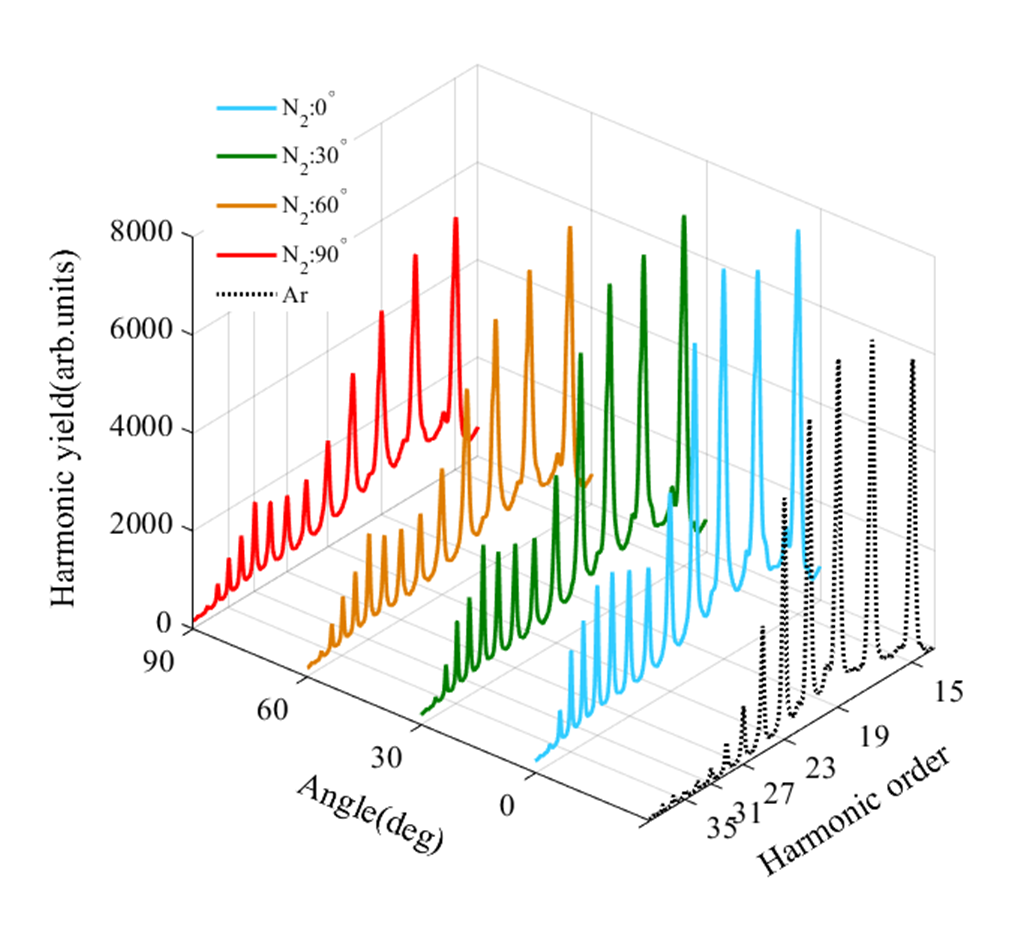}}
\caption{\label{n2angle}(Color online) High harmonic spectra of N$_2$ measured at different alignment angles. The delay between the pump and probe pulses is 4.1 ps. The dotted line shows the high harmonic spectrum of reference atom Ar.}
\end{figure}

\begin{figure}[!t]
\centerline{
\includegraphics[width=8cm]{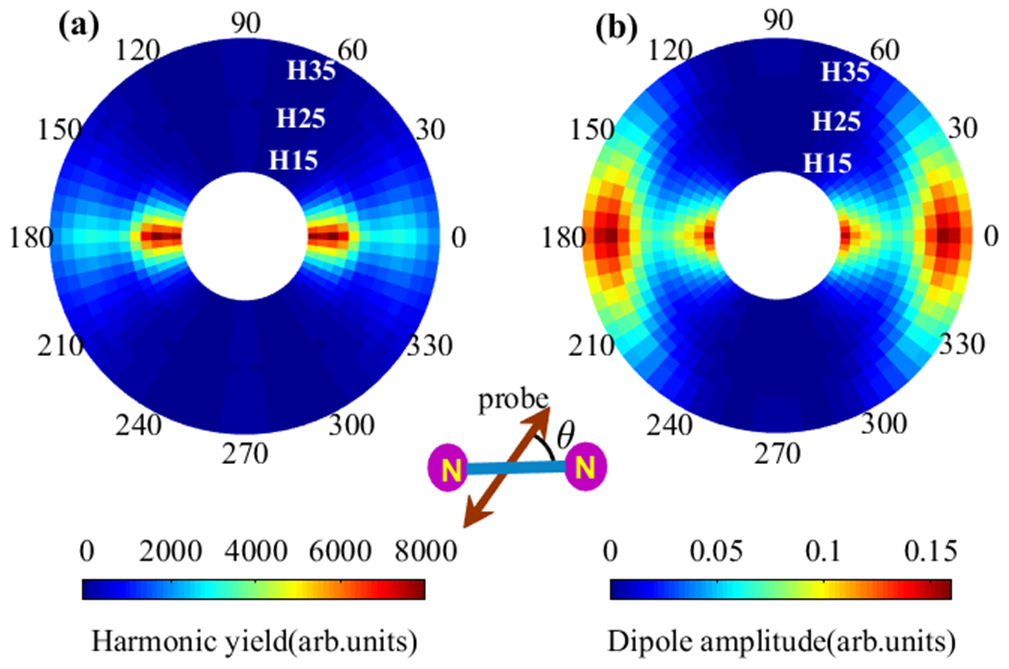}}
\caption{\label{dipole}(Color online) Angular distribution of the high harmonics generated from N$_2$. (a), Polar plot of the deconvoluted high harmonic spectra of N$_2$. (b),  The amplitude of recombination dipole as a function of alignment angle and harmonic order. The recombination amplitude is obtained from the measured high harmonic spectra of N$_2$ with those of reference atom Ar.}
\end{figure}

Next the high harmonics generated at 4.1 ps for different alignment angles are measured. Figure \ref{n2angle} shows the high harmonic spectra of N$_2$ measured for different alignment angles (solid lines) and the high harmonic spectrum of reference atom Ar (dotted line). Recall that the high-harmonic signal measured in the macroscopic molecular ensemble in the laboratory frame is a convolution of the single molecular signal with the alignment distribution. Deconvolution is carried out as in Refs. \cite{r40,r41}. Figure \ref{dipole}(a) shows the deconvolution signal for the harmonics from H15 to H37, which present a prolate distribution. As in Ref. \cite{r8}, the experimental data has been extrapolated up to 360$^\circ$ by imposing the assumed symmetry of the highest occupied molecular orbital (HOMO) of N$_2$. Then the amplitude of the recombination dipole is obtained from the high harmonic spectra of N$_2$ divided by that of reference atom Ar. The obtained recombination dipole amplitude is shown in Fig. \ref{dipole}(b) as a function of harmonic order and angle, which corresponds to the diffraction pattern of the molecular orbital. It is worth emphasizing that the harmonic phase is not measured in our experiment, i.e., only the amplitude of the recombination dipole is obtained.

\begin{figure}[!t]
\centerline{
\includegraphics[width=8cm]{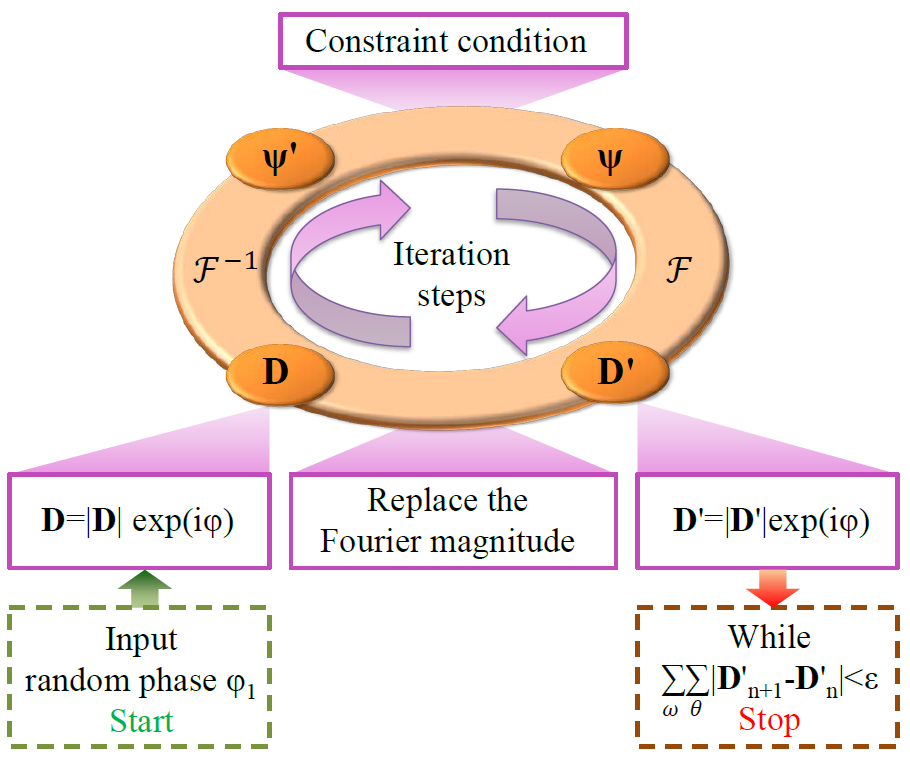}}
\caption{\label{loop}(Color online) Schematic diagram of the guided iterative algorithm for molecular orbital reconstruction.}
\end{figure}
\begin{figure}[!b]
\centerline{
\includegraphics[width=8.5cm]{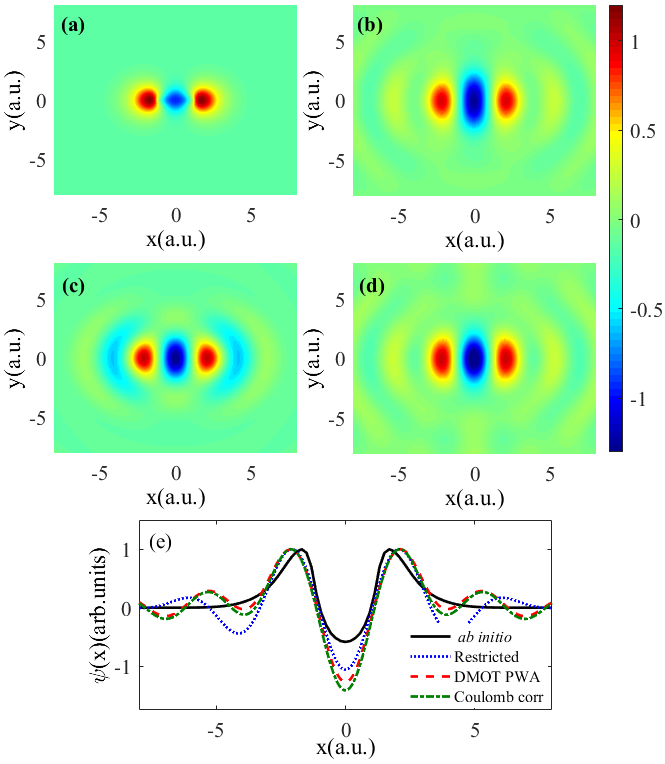}}
\caption{\label{n2}(Color online) (a), Calculated molecular orbital of N$_2$ with the \emph{ab initio} method. The orbital is projected to the polarization plane. (b), Reconstructed orbital from the experimental data shown in Fig. \ref{dipole} with the DMOT method. (c), Restricted orbital simulated with the limited frequency range according to the experimental conditions. (d), Coulomb corrected orbital reconstructed from the experimental data shown in Fig. \ref{dipole}. (e), Cuts along the internuclear axis for the molecular orbitals.}
\end{figure}
\begin{figure}[!t]
\centerline{
\includegraphics[width=8cm]{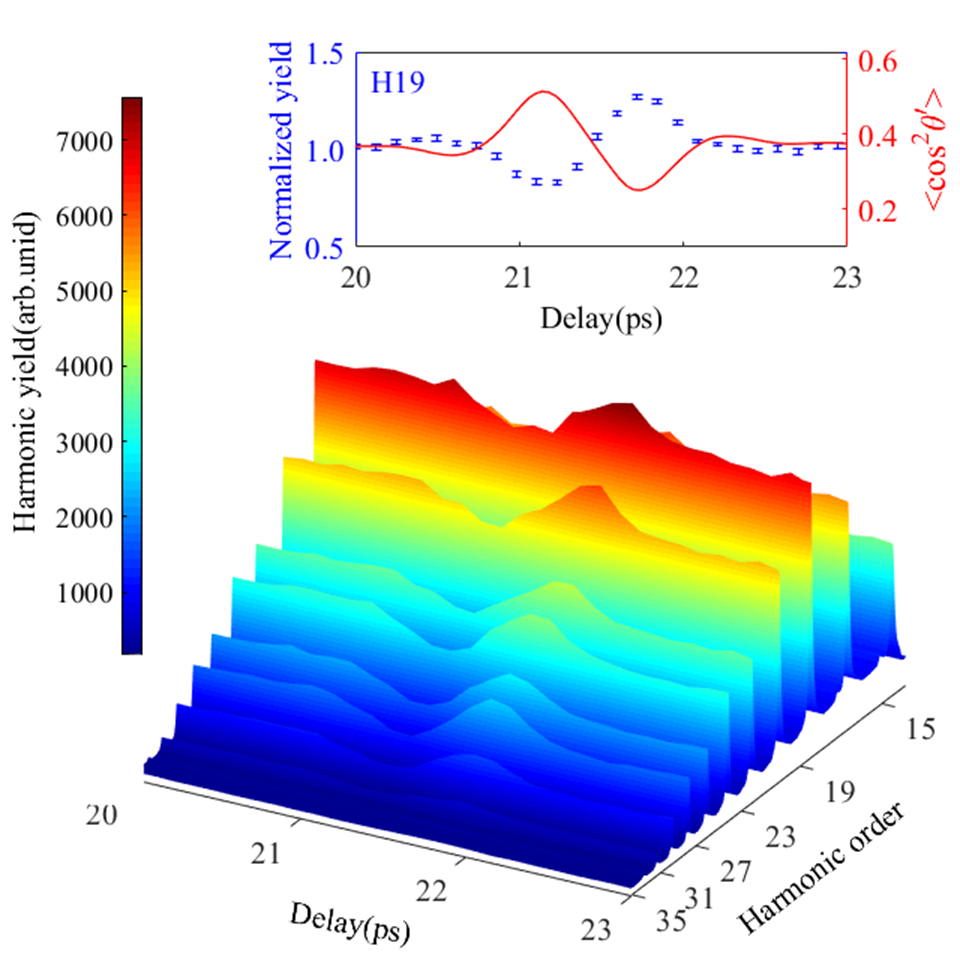}}
\caption{\label{co2delay}(Color online) High harmonic spectra of CO$_2$ molecules measured as a function of the pump-probe delay. The pump and probe pulses have parallel polarizations. The solid line in the inset displays the calculated temporal evolution of the alignment degree $\langle\cos^2{\theta}^\prime\rangle$. The dots display normalized HHG yield of the 19th harmonic.}
\end{figure}
\begin{figure}[!b]
\centerline{
\includegraphics[width=8cm]{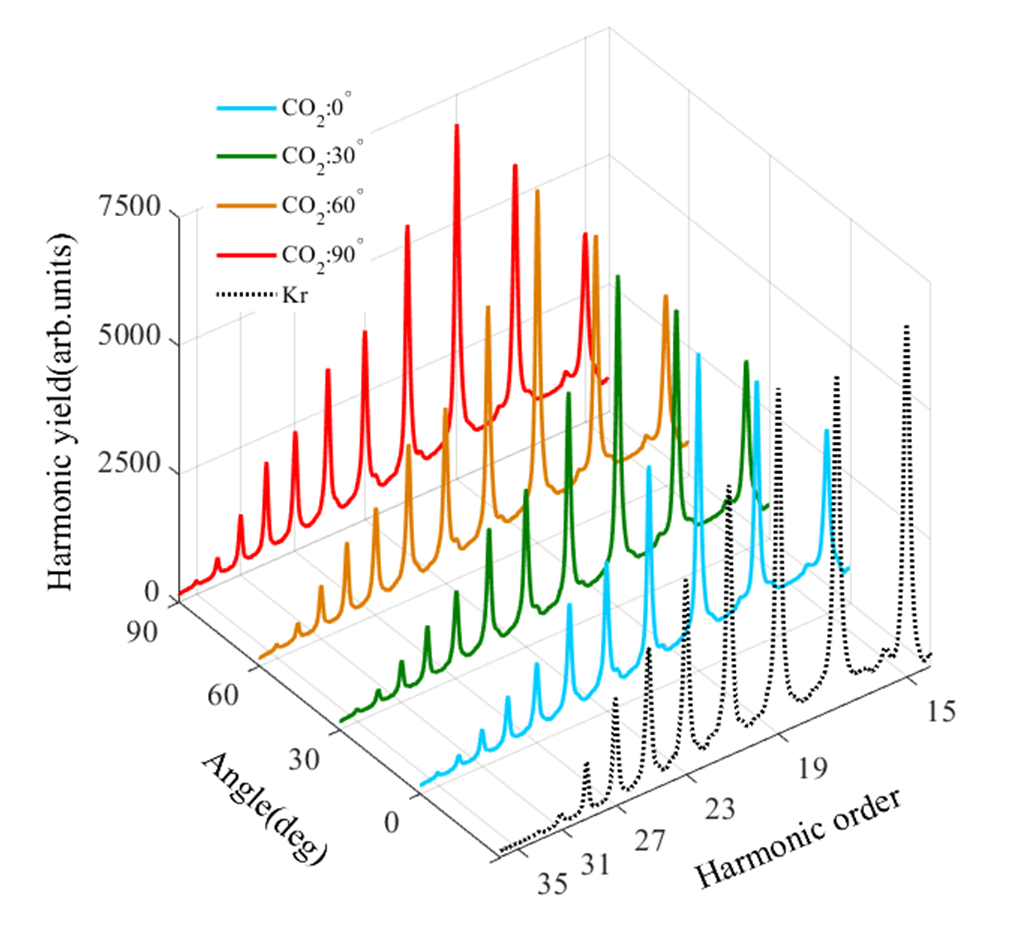}}
\caption{\label{co2angle}(Color online) High harmonic spectra of CO$_2$ measured at different alignment angles. The delay between the pump and probe pulses is 21.1 ps. The dotted line shows the high harmonic spectra of reference atom Kr.}
\end{figure}
\begin{figure}[!t]
\centerline{
\includegraphics[width=8cm]{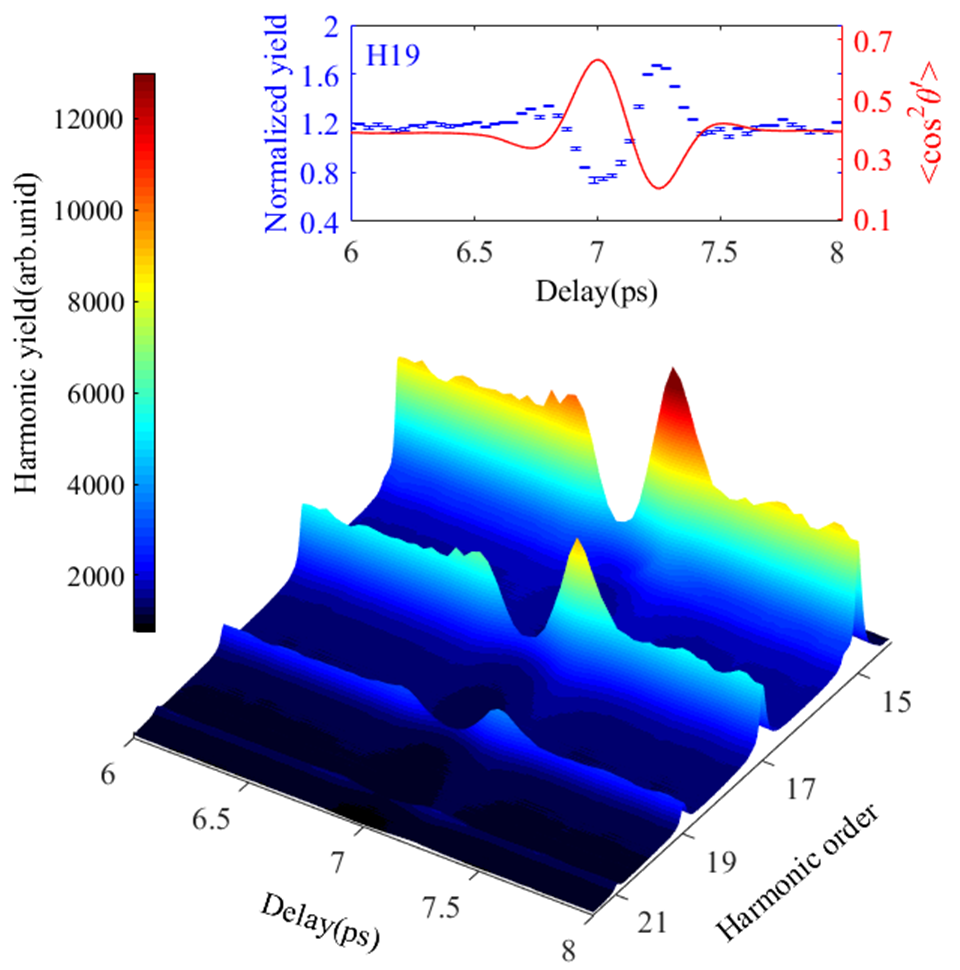}}
\caption{\label{c2h2delay}(Color online) High harmonic spectra of C$_2$H$_2$ molecules measured as a function of the pump-probe delay. The pump and probe pulses have parallel polarizations. The solid line in the inset displays the calculated temporal evolution of the alignment degree $\langle\cos^2{\theta}^\prime\rangle$. The dots display normalized HHG yield of the 19th harmonic.}
\end{figure}
\begin{figure}[!b]
\centerline{
\includegraphics[width=8cm]{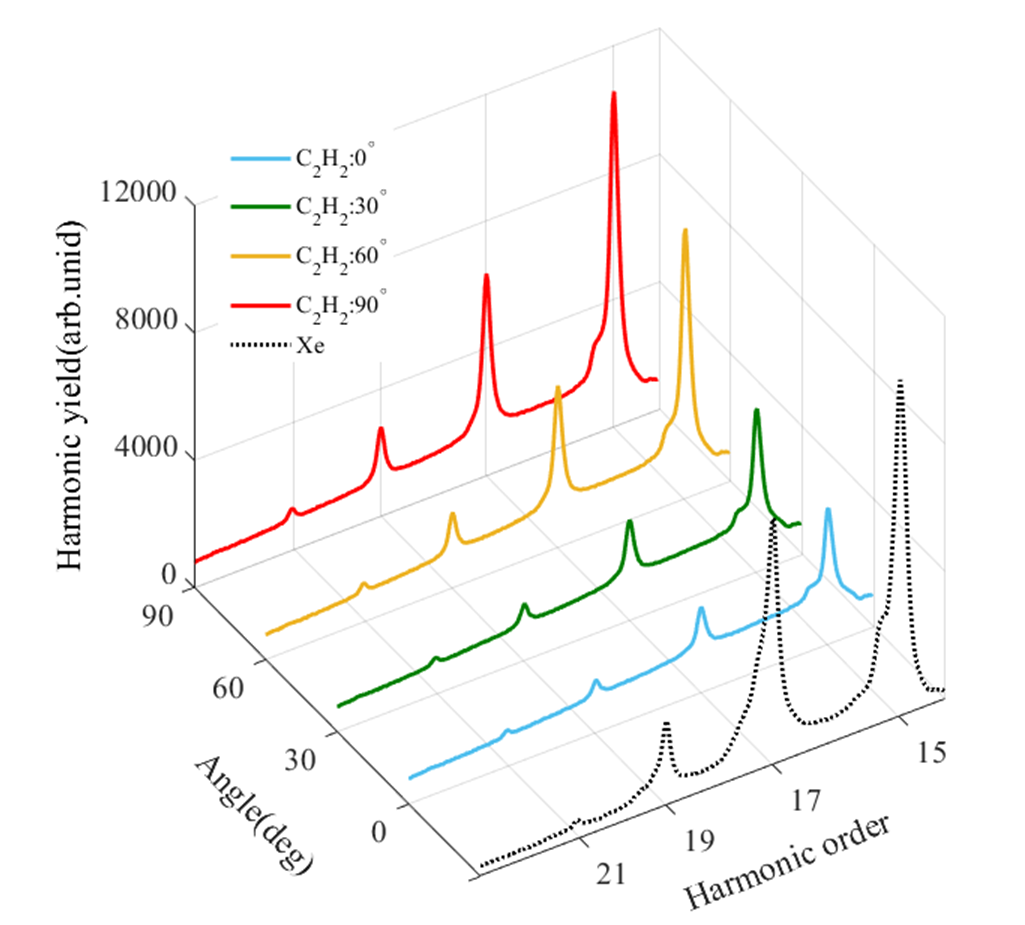}}
\caption{\label{c2h2angle}(Color online) High harmonic spectra of C$_2$H$_2$ measured with different alignment angles. The delay between the pump and probe pulses is 7.04 ps. The dotted line shows the high harmonic spectra of reference atom Xe.}
\end{figure}

\begin{figure*}
\centerline{
\includegraphics[width=16cm]{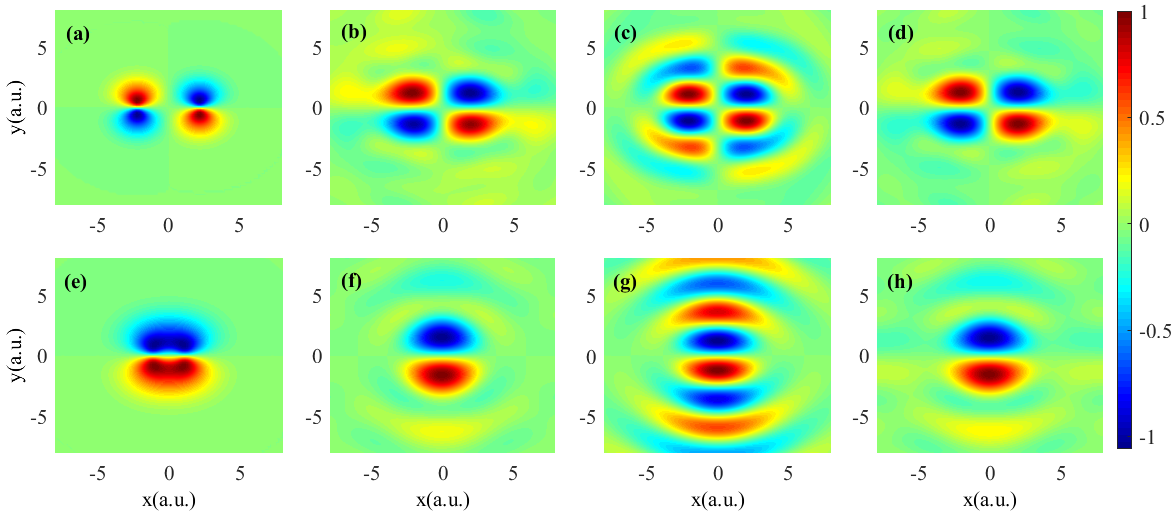}}
\caption{\label{co2c2h2}(Color online) Reconstructed HOMO of CO$_2$ (a-d) and C$_2$H$_2$ (e-h). The orbital is reconstructed with the same procedure as N$_2$ shown in Fig. \ref{n2}. For CO$_2$, the laser intensity is changed to about $1.9\times10^{14}$ W/cm$^2$ and the reference atom of Kr is adopted. For C$_2$H$_2$, the laser intensity is changed to about $1.2\times10^{14}$ W/cm$^2$ and the reference atom of Xe is adopted.}
\end{figure*}

The orbital is retrieved from the amplitude of molecular transition dipole shown in Fig. \ref{dipole}(b). For this purpose, an iterative retrieval algorithm is developed. For convenience, we define $D=|\tilde{D}|=|\tilde{d}|/k$. The procedure of the iterative retrieval algorithm is illustrated in Fig. \ref{loop}. In brief, the iteration is started with the experimental values $D^{expt}$ and random phase values $\varphi_1$, $\tilde{D}_1=D^{expt}$exp$(i\varphi_1)$, where the subscript denotes the \emph{n}th iteration. Then, in each loop, we perform the following four steps: (1) obtain a molecular orbital $\Psi^\prime_n$ by inverse Fourier transform of $\tilde{D}_n$; (2) obtain the revised orbital $\Psi_n$ by applying the constraint on $\Psi^\prime_n$ following the ER (or HIO) algorithm \cite{rdiff,rmiao}; (3) obtain a new complex amplitude $\tilde{D}^\prime_n$ by Fourier transform of the orbital $\Psi_n$; (4) replace the amplitude of $\tilde{D}^\prime_n$ with the experimental values of amplitude as $\tilde{D}_{(n+1)}=D^{expt}$exp$(i\varphi_n)$ and return to (1) to start the next iteration. The iteration can be stopped while the results are convergent. Guided algorithm is applied to speed up the convergence \cite{rghio} (see more details in the Appendix A).

The reconstructed HOMO of N$_2$ is shown in Fig. \ref{n2}(b) and the orbital calculated with the \emph{ab initio} method \cite{r43} is shown in Fig. \ref{n2}(a) for comparison. One can see that the retrieved orbital faithfully reproduces the main characteristics of the \emph{ab initio} orbital, namely three alternating positive and negative lobes and two nodal planes passing through each nucleus. The differences between the reconstructed orbital (see Fig. \ref{n2}(b)) and the \emph{ab initio} orbital (see Fig. \ref{n2}(a)) are the size of the lobes and the separation between the nodal planes (i.e., the internuclear distance). This can be seen more clearly from the line profiles along the internuclear axis ($y$=0) of the orbitals. As shown in Fig. \ref{n2}(e), the internuclear distance of the reconstructed orbital is 2.16 a.u. (the red dashed line), which is very close to but slightly larger than that of the \emph{ab initio} orbital (2.06 a.u., the black solid line). This difference can be attributed to the limited spectral range in our experiment \cite{r11}, which spans from H15 to H37 and the lower and higher order harmonics are not measured in our experimental conditions. To mimic this spectral limitation, we simulate the restricted recombination dipoles from the \emph{ab initio} orbital over the spectral range and alignment angles according to our experiment. Then the restricted orbital is obtained by the inverse Fourier transform of the restricted dipoles. Figure \ref{n2}(c) shows the restricted orbital and the line profile is shown in Fig. \ref{n2}(e) (the blue dotted line). One can see that the reconstructed orbital using the experimental data agrees very well with the restricted orbital, with only minor distinctions of the center lobe size.

To further improve the quality of the retrieved orbital, we consider taking account of the correction of the Columbic potential. Recall that the original MOT theory relies on the strong-field approximation, where the freed electron wavepacket after tunneling is solely determined by the laser field and the influence of the Columbic potential is ignored. The continuum electron wave functions are approximated by plane waves (PWs) \cite{r8}. Such an assumption has led to the beautiful formula relating the transition dipole to be Fourier transform of the molecular orbital. However, considering that high harmonic photon energy is less than 100 eV in MOT experiments, the Coulomb potential can play a role in HHG and better description of the continuum electron wavepacket is desired \cite{r17,r18}. In this work, we exploit a correction by improving the expression of the continuum electron wavepacket. Instead of the PWs, we describe the continuum electron wavepacket by a two-center Coulomb wave function \cite{r44}, which is the solution of the scattering state of the two-body Coulomb problem and therefore describes the continuum wavepacket and recombination dipole more faithfully (see Appendix B).
The reconstructed orbital with Coulomb correction is shown in Fig. \ref{n2}(d). One can see that the center lobe is compressed in the longitudinal direction, agreeing with the restricted \emph{ab initio} orbital better than that of the reconstructed orbital with PWs. Except the correction of the center lobe, the improvement of the Coulomb correction is not remarkable compared to DMOT. One can even see some oscillations at $|\mathbf{r}|>2$ a.u. with the amplitude of $20\%$ compared to the central peak. It is because that the limited spectral range dominates the quality of the retrieved orbital in our experimental conditions. To clarify the influence of the spectral range, we have theoretical calculated the high harmonics of N$_2$ and its reference atom Ar. It indicates that if the harmonic cutoff can be extended to H59, the retrieved orbital with DMOT becomes almost coincides with the \emph{ab initio} orbital except some oscillations with the amplitude of $5\%$ of the center peak. Moreover, the oscillations can be removed by using Coulomb correction DMOT \cite{r44}.

Coherent diffractive imaging is barely dependent on the sample structure. DMOT scheme inherits this property and is in principle straightforward for extending to other molecules. The issues that require special attentions are the molecular alignment and the possible multielectron contributions. To demonstrate the extension, we carried out the experiments with CO$_2$ and C$_2$H$_2$. Note that these molecules have $\pi_g$ and $\pi_u$ HOMO, different from the HOMO of N$_2$. Moreover, these molecules can be impulsively aligned and are also suitable for the comparison with previous investigations \cite{r10}. The experiment is performed following the same procedure discussed above, but with lower probe laser intensities ($1.9\times10^{14}$, $1.2\times10^{14}$ W/cm$^2$ for CO$_2$ and C$_2$H$_2$, respectively) and different reference atoms (Kr and Xe, respectively). Figure \ref{co2delay} shows the delay dependent high harmonic spectra and Fig. \ref{co2angle} shows the high harmonic spectra at different alignment angles for CO$_2$. Figure \ref{c2h2delay} and \ref{c2h2angle} show the results for C$_2$H$_2$.
It should be noted that previous experiments have revealed the multielectron effects for CO$_2$ \cite{r15,r45}. It was shown that the contribution of lower lying orbital (e.g., HOMO-2 of CO$_2$) is comparable to that of HOMO for the high harmonics in the cutoff region, which leads to a minimum in the harmonic spectra \cite{r15,r45}. This minimum depends on the laser intensity and is around H29 at about $1.8\times10^{14}$ W/cm$^2$ for the 800-nm laser. However, in our experiment, one can see from Fig. \ref{co2angle} that the yield of the harmonics higher than H29 rapidly decreases in the cutoff region. Therefore, the spectral minimum is not visible for all the alignment angles from 0 to 90 degrees. This result indicates that the contribution of HOMO is dominant for the plateau high harmonics in our experimental condition. Note that the ionization energies of HOMO-1 (18.1eV) and other lower lying orbitals of C$_2$H$_2$ are much higher than that of the HOMO (11.4 eV). Therefore, the tunneling ionization and HHG are expected to be dominant by the contribution of HOMO for C$_2$H$_2$.

Figure \ref{co2c2h2} shows the reconstructed orbitals of CO$_2$ and C$_2$H$_2$ molecules. One can see that the retrieved orbitals (Fig. \ref{co2c2h2}(b) and (f)) faithfully reproduce the \emph{ab initio} orbitals, with only subtle difference of the lobe size. The correction of the Coulomb potential can finely improve the size and lead to a better agreement as shown in Fig. \ref{co2c2h2}(d) and (h).
\noindent \section{Conclusions}
In conclusion, the combination of diffractive imaging with MOT brings a new perspective on the imaging of molecule orbitals. With this DMOT scheme, the molecular orbital can be retrieved solely from the amplitude of the HHG without measuring the phase. Moreover, DMOT inherits the advances of diffractive imaging and hence is suitable to be extended to more complex molecules other than N$_2$, CO$_2$ and C$_2$H$_2$.

Even though only the stationary orbital is obtained in our experiment, the HHG intrinsically contains the subfemtosecond timing information because HHG is produced by the electron recollision in a subcycle of the laser field \cite{r46,r47}. Therefore, this high temporal and spatial resolution holds the potential to make the ultrafast movies of molecular dynamics (i.e., time-dependent orbitals), which is one intriguing goal of ultrafast optics and physical chemistry \cite{r48}. To this end, the procedures of MOT should be simplified so as to efficiently capture the snapshot. The DMOT scheme circumvents the hurdle of phase measurement, making a substantial step towards this goal. On the other hand, some challenges have to be addressed for imaging the time-dependent orbitals. For instance, the time-dependent orbital could be a complex-valued wavefunction, which make it very difficult to obtain the convergent solution with the iterative algorithm. In addition, lots of efforts are needed to concentrate on the complex dipole and MOT theory.

\noindent \section*{ACKNOWLEDGMENT}
We gratefully acknowledge M. Lein for valuable discussions and reading the manuscript.
This work was supported by the National Natural Science Foundation of China under Grants No. 11422435, 11234004, 61275126 and 11404123. Numerical simulations presented in this paper were carried out using the High Performance Computing experimental testbed in SCTS/CGCL (see http://grid.hust.edu.cn).

\appendix
\section{Guided algorithm for molecular orbital reconstruction}
To explain the algorithm for molecular orbital reconstruction, let us briefly outline the theory for high harmonic generation (HHG). Based on the laser-induced recollision model, the HHG process is understood by the ionization, acceleration and recombination steps. Therefore, the high harmonic radiation can be factorized as \cite{r8,r47,r49}
\begin{equation}
\tilde{\mathbf{E}}(\omega,\theta)\propto \alpha_{ion}(\mathbf{k})\alpha_{acc}(\mathbf{k})\tilde{\mathbf{d}}(\mathbf{k}),
\end{equation}
where $\alpha_{ion}(\mathbf{k})$, $\alpha_{acc}(\mathbf{k})$ represent the ionization and acceleration amplitudes, respectively. $\mathbf{k}$ is the momentum of the returning electron. $\tilde{\mathbf{d}}(\mathbf{k})$ is the recombination dipole moment. As discussed in \cite{r8,r47}, $\alpha_{ion}(\mathbf{k})$ and $\alpha_{acc}(\mathbf{k})$ depend insensitively on the target structure. Hence, $\tilde{\mathbf{d}}(\mathbf{k})$ can be extracted from the experimental measurement by dividing the detected high harmonic radiation from the target molecule by that from a reference atom with the comparable ionization energy.
The recombination dipole can be written in both the length form and the velocity form \cite{r34,r47}. In this work, we adopt the velocity form, but the length form can also be adopted as in \cite{r8}. Following \cite{r34,r47}, the returning continuum electron wavepackets are approximated as plane waves and the recombination dipole moment in the velocity form reads:
\begin{align}
\tilde{\mathbf{d}}(\mathbf{k})&=\mathbf{k}\int\Psi_0(\mathbf{r})e^{i\mathbf{k}\cdot\mathbf{r}}d\mathbf{r}\\
\label{FT} &=\mathbf{k}\mathcal{F}[\Psi_0(\mathbf{r})]
\end{align}
where $\Psi_0(\mathbf{r})$ is the molecular orbital. It is shown in Eq. (\ref{FT}) that, the molecular orbital $\Psi_0(\mathbf{r})$ and the measurable quantity $\tilde{\mathbf{d}}(\mathbf{k})/\mathbf{k}$ are Fourier transform pairs. If both the phase and amplitude of HHG can be measured as functions of alignment angle and harmonic order, the full information of $\tilde{\mathbf{d}}(\mathbf{k})/\mathbf{k}$ can be obtained and then the molecular orbital can be directly imaged by inverse Fourier transform
\begin{equation}
\Psi_0(\mathbf{r})=\mathcal{F}^{-1}[\tilde{\mathbf{d}}(\mathbf{k})/\mathbf{k}]
\end{equation}
Unfortunately, only the intensity of the high harmonic spectra can be directly measured with the spectrometer, i.e., $S(\omega,\theta)=|\tilde{\mathbf{E}}(\omega,\theta)|^2$. Therefore, only the amplitude of the recombination dipole is determined. As discussed in Sec. II. A, the high harmonic spectra are sampled with an interval several times finer than the Nyquist interval. Therefore, the phase information is in principle encoded in the oversampled high harmonic spectra, enabling to reconstruct the molecular orbital with the high harmonic intensity without measuring the phase. To this end, we have developed the guided iterative algorithm to retrieve the molecular orbital. The reconstruction is carried out as follows.

In the momentum (i.e., $\mathbf{k}$) space, for convenience, we define
\begin{equation}
{D}(\mathbf{k})=|\tilde{D}(\mathbf{k})|=\frac{|\tilde{d}(\mathbf{k})|}{k}
\end{equation}
The iterative reconstruction procedure in our diffractive molecular orbital tomography (DMOT) goes as follows.
As shown in Fig. \ref{loop}, we start the iterations with the initial complex amplitude $\tilde{D}_1(\mathbf{k})=D^{expt}(\mathbf{k})$exp${[i\varphi_1(\mathbf{k})]}$ by employing the experimental amplitude $D^{expt}(\mathbf{k})$ and the random phase  $\varphi_1(\mathbf{k})$. Then, in each loop of iteration, the algorithm consists of four steps \cite{rghio}:

   (i) Obtain a molecular orbital $\Psi^\prime_n(\mathbf{r})$ in the coordinate $(\mathbf{r})$ space by the inverse Fourier transform of $\tilde{D}_n(\mathbf{k})$, i.e., $\Psi^\prime_n(\mathbf{r})=\mathcal{F}^{-1}[\tilde{D}_n(\mathbf{k})]$. Here, $n$ denotes the \emph{n}th iteration. Note that one can adopt the fast Fourier transform (FFT) and the fast inverse Fourier transform (FFT$^{-1}$) algorithm to speed up the calculation.

   (ii) Revise the molecular orbital with the constraint condition. The molecular orbital is revised according to the ER algorithm (Eq. (\ref{ER})) or the HIO algorithm (Eq. \ref{HIO}) \cite{rdiff,rmiao} (In practice, the ER algorithm is combined with HIO algorithm to improve its performance.) and the symmetries of the highest occupied molecular orbitals (HOMOs) (i.e., $\sigma_g$ for N$_2$, $\pi_g$ for CO$_2$ and $\pi_u$ for C$_2$H$_2$). Then, we obtain the revised molecular orbital $\Psi_n(\mathbf{r})$ with

   ER algorithm:\\
\begin{equation}\label{ER}
\Psi_n(\mathbf{r})=\left\{
\begin{array}{rl}
&\Psi^\prime_n(\mathbf{r})     \hspace{2.5cm}|x| \textrm{and} |y|\leq5 \textrm{a.u. }          \\
&0 \hspace{3.2cm}\textrm{otherwise}
\end{array}
\right.
\end{equation}

   HIO algorithm:\\
\begin{equation}\label{HIO}
\Psi_n(\mathbf{r})=\left\{
\begin{array}{rl}
&\Psi^\prime_n(\mathbf{r})     \hspace{2.5cm}|x|\textrm{and}|y|\leq5 \textrm{a.u.}           \\
&\Psi^\prime_{n-1}(\mathbf{r})-\gamma\Psi^\prime_n(\mathbf{r}) \hspace{0.6cm}\textrm{otherwise}
\end{array}
\right.
\end{equation}
where $x=r\cos(\theta)$, $y=r\sin(\theta)$ and $\gamma=0.9$ in our reconstruction procedure.

   (iii) Obtain a new complex amplitude by the Fourier transform of $\Psi_n(\mathbf{r})$, i.e., $\tilde{D}^\prime_n(\mathbf{k})=\mathcal{F}[\Psi_n(\mathbf{r})]$.

   (iv) Replace the amplitude of $\tilde{D}^\prime_n(\mathbf{k})$ with ${D^{expt}}(\mathbf{k})$ and obtain a new complex amplitude to start the $(n+1)$th iteration, $\tilde{D}_{(n+1)}(\mathbf{k})=D^{expt}(\mathbf{k})$exp$[i\varphi_n(\mathbf{k})]$.
We define an error function $erf F$ to evaluate the quality of the molecular orbital by comparing the retrieved recombination dipole with the experimentally measured dipole
\begin{equation}
erf F=\frac{1}{n_\omega}\frac{1}{n_\theta}\sum_\omega\sum_\theta
\frac{\big||D^{expt}(\omega,\theta)|-|\tilde{D}^\prime_n(\omega,\theta)|\big|}
{|D^{expt}(\omega,\theta)|}
\end{equation}
With increasing the iteration loops, $erf F$ gradually reduces and finally convergent molecular orbital can be obtained. To speed up convergence, we use a guided iterative algorithm \cite{rghio}: we start the iterative procedure for one hundred independent runs with one hundred initial complex amplitudes. After repeating the steps (i) to (iv) 200 times, we stop the iterative procedure and define them to be the first generation (G1). Then, we select the molecular orbital with the smallest $erf F$ to be the ``favorable gene", $\Psi_{gene}$. To start the second generation (G2), we use the ``favorable gene" to guide the molecular orbitals $\Psi_{G1}$, by taking the square root of the product of $\Psi_{gene}$ and $\Psi_{G1}$
\begin{equation}
\Psi=\sqrt{\Psi_{gene}\times\Psi_{G1}}
\end{equation}
We, therefore, calculate the second generation by repeating the iterative loops again. After 200 iterations, we obtain one hundred molecular orbitals in the second generation G2 and their $erf F$. By comparing the $erf F$ of G2 with that of G1, one can see that the guiding algorithm has passed the ``favorable gene" to the succeeding generations, thus the $erf F$ of G2 is reduced. Using the same guided algorithm, we calculate the third generation G3 and find that all of the results are convergent for N$_2$. Hence, the molecular orbital of N$_2$ can be uniquely and successfully reconstructed after 3 generations. The molecular orbitals of CO$_2$ and C$_2$H$_2$ can be also reconstructed by using the same guided iterative algorithm within 3-5 generations.
\section{DMOT with Coulomb correction}
To exploit a correction of the Coulomb effect in DMOT, a more accurate way is adopted to express the continuum electron wavepacket as the two-center Coulomb (TCC) wave function with outgoing boundary conditions \cite{r44}. The TCC can be written as
\begin{equation}
\Psi_{\mathbf{k}}^{TCC}(\mathbf{r})=\frac{1}{(2\pi)^{3/2}}C(\mathbf{k},\mathbf{r}_1)C(\mathbf{k},\mathbf{r}_2)e^{i\mathbf{k}\cdot\mathbf{r}}
\end{equation}
with
\begin{equation}
C(\mathbf{k},\mathbf{r})=\Gamma(1-i\nu)_1F_1[i\nu,1,i(kr-\mathbf{k}\cdot\mathbf{r})]e^{\pi\nu/2}
\end{equation}
where $\mathbf{r}_1=\mathbf{r}+\mathbf{R}/2$ and $\mathbf{r}_2=\mathbf{r}-\mathbf{R}/2$. Here, $\mathbf{R}$ is the internuclear distance. $F_1$ is the confluent hypergeometric function. $\nu=Z/k$ is the Sommerfeld parameter, where $Z$ is the effective ion charge. We set $Z=1/2$ for each ion to match the condition that the molecular ion acts on the recolliding electron with the effective charge of $+1$ asymptotically. $\Psi_{\mathbf{k}}^{TCC}(\mathbf{r})$ can be expanded in the momentum $\mathbf{k}$ space by performing the transformation
\begin{equation}
\Psi_{\mathbf{k}}^{TCC}(\mathbf{k^\prime})=\int\Psi_{\mathbf{k}}^{TCC}(\mathbf{r})e^{-i\mathbf{k}^\prime\cdot\mathbf{r}}d\mathbf{r}
\end{equation}
In this way, the recombination dipole can be rewritten as $\tilde{\mathbf{d}}(\omega,\theta)=\mathbf{k}\mathbb{S}\int\Psi_{0}(\mathbf{r})$exp$(-\mathbf{k^\prime\cdot\mathbf{r}})=\mathbf{k}\mathbb{S}\mathcal{F}(\Psi_{0}\mathbf{r})$, where $\mathbb{S}(\mathbf{k},\mathbf{k^\prime})=\Psi_{\mathbf{k}}^{TCC}(\mathbf{k^\prime})$ is the transformation matrix \cite{r44}. Our numerical simulation indicates that the matrix $\mathbb{S}$ is invertible for the continuum electron wavepackets with energy from 0 to 3.17 Up. Then the molecular orbital is still related to the Fourier transform of the measurable quantity as $\Psi_0(\mathbf{r})=\mathcal{F}^{-1}[\mathbb{S}^{-1}\tilde{D}(\mathbf{k})]$, and $\Psi_0(\mathbf{r})$ can be retrieved with the guided iterative algorithm with $\mathbb{S}^{-1}$ being calculated in advance.

\end{document}